\documentclass[
 reprint,
 amsmath,amssymb,prd,
 aps,numerical,nofootinbib
]{revtex4-1}
\usepackage{graphicx}
\usepackage{dcolumn}
\usepackage{bm}
\usepackage{natbib}
\usepackage{stackrel}
\usepackage{mathtools}
\usepackage[english]{babel}
\usepackage[utf8]{inputenc}
\usepackage{amsmath, amssymb}
\usepackage{slashed}		
\usepackage{braket}

\usepackage[pdfencoding=auto, psdextra]{hyperref}
\usepackage{epsfig,epsf}

\begin{document}

\title{B \texorpdfstring{$\rightarrow f_2(1270)$}{ to f2} form factors with light-cone sum rules }

\author{M. Emmerich}
\email{maximilian.emmerich@ur.de}
\author{M. Strohmaier}%
 \email{matthias.strohmaier@ur.de}
\author{A. Schäfer}%
 \email{andreas.schaefer@ur.de}
 \affiliation{Institute for Theoretical Physics, University of Regensburg, D-93053 Regensburg, Germany}

\date{\today}

\begin{abstract}
We construct the quark-antiquark  chiral odd distribution amplitudes including twist-four mass contributions for tensor mesons. 
We also give quark-antiquark-gluon distribution amplitudes, where we calculate the input parameters with QCD sum rules. 
With the help of equations of motion we determine the twist-three and twist-four distribution amplitudes including SU(3) breaking terms.
We use QCD light-come sum rules to derive the form factors for the decay B $\rightarrow f_2(1270)$ with vector, axial-vector and tensor currents.
We also give the $q^2$ dependence of the form factors.

\end{abstract}

\maketitle

\section{Introduction}

The decay channel $B\rightarrow meson+\ell\ell$  with mesons which are light compared to the B 
is an especially promising one in the search for New Physics. This can be illustrated, e.g., by 
the recent discussion about angular distributions in $B \rightarrow K^{*0}\mu^+\mu^-$
\cite{Aaij:2015oid,ATLAS:2017dlm,Sirunyan:2017dhj}. Decays into $f_2+\ell\ell$ have the advantage that
the polarization of the final tensor meson provides additional sensitivity to search for 
deviations from the helicity structure of the electroweak interaction. 
(For some general introduction see the mini-reviews by A. Gritsan (page 1252-1255 in the 2017 online update)
and P. Eerola, M. Kreps and Y. Kwon (page 1137-1149) in \cite{Patrignani:2016xqp} and references given there.)    
In fact, it was demonstrated 
by BELLE in a recent measurement of the transition form factor 
$\gamma^* \gamma \rightarrow f_2(1270)$ at large momentum transfers 
that already with the existing detectors relevant polarization sensitive data can be 
obtained \cite{Masuda:2015yoh}. 
The uncertainty of the standard model predictions, with which any experimental result would have to be compared,
is dominated by QCD uncertainties, namely by the uncertainties of the $B\rightarrow f_2(1270)$ decay form 
factors, which are the topic of this contribution.

Tensor mesons have already been the topic of earlier work. In Ref. \cite{1007.3541v1} the chiral even and odd distribution amplitudes (DAs) were constructed and the decay constants were calculated,
while in Ref. \cite{Braun2016} the chiral even DAs including meson mass corrections and three-particle twist three DAs were studied.  The present contribution is largely based on that work.
The definitions of the B to tensor meson form factors can be found in \cite{Hatanaka:2009gb,Hatanaka:2009sj,1008.5326v2}. 
There are a few studies of the B to $f_2(1270)$ decay, for example using a perturbative QCD approach \cite{1008.5326v2} or using light-cone sum rules \cite{1010.2944v3,Wang:2010tz}.

In this paper we calculate the form factors for the B meson decaying into the tensor meson $f_2(1270)$ by using the framework of  light cone sum rules (LCSR) \cite{Balitsky:1986st,Balitsky:1989ry,Chernyak:1990ag}.
We give for the first time the chiral odd quark-antiquark DAs, including higher twist contributions and meson mass corrections.  We also construct new three particle quark-antiquark-gluon DAs  with tensor structure.
With the help of equations of motion (EOM) we can represent the higher twist DAs in terms of lower twist DAs including SU(3) breaking terms for the first time.
We determine quark-gluon coupling constants appearing in the three particle DAs using QCD sum rules.
In doing so we assume that $f_2(1270)$ is a pure nonstrange isospin singlet state  $1/\sqrt{2}(\bar{u}u+\bar{d}d)$ and $f_2'(1525)$ is a pure strange state $\bar{s} s$   which is equivalent to assuming a vanishing mixing angle \cite{hep-ph/0010342v2,1674-1137-38-9-090001}. 

The paper is organized as follows. In section II we give the form factors and the related LCSR expressions. Section III contains the numerical analysis of the sum rules and our results. In the appendix we define the leading and higher twist DAs of the tensor mesons.

\section{Form factors and light cone sum rules}
We define the semileptonic $B \rightarrow f_2(1270)$ form factors by  \cite{Hatanaka:2009gb, Hatanaka:2009sj,1008.5326v2}%
\begin{align}
&\bra{f_2^\lambda (P)}  \bar{u}(0)\gamma_\mu b(0) \ket{B(P')}= \nonumber \\
& \qquad \quad \frac{2}{m_B+m_{f_2}} \epsilon_{\mu \nu \alpha \beta} e_{(\lambda)*}^{\nu} P'^\alpha P^\beta \tilde{V}(q^2), \label{fofa1}\\
&\bra{f_2^\lambda(P)} \bar{u}(0) \gamma_\mu \gamma_5 b(0) \ket{B(P')}=\nonumber \\
& \qquad \quad i (m_B+m_{f_2}) e^{(\lambda)* }_\mu \tilde{A}_1(q^2) \nonumber \\
&\qquad \quad-i \frac{e^{(\lambda)*} \cdot P'}{m_B+m_{f_2}}  (P'+P)_\mu \tilde{A}_2(q^2)\nonumber \\
&\qquad \quad-2 i m_{f_2} \frac{e^{(\lambda)*} \cdot P'}{q^2} q_\mu [\tilde{A}_3(q^2)-\tilde{A}_0(q^2)] ,\\
&\bra{f_2^\lambda(P)} \bar{u}(0) \sigma_{\mu \nu} \gamma_5 b(0) \ket{B(P')}=\nonumber\\
& \qquad \quad \tilde{A}(q^2) \left[ e^{(\lambda)*}_\mu (P+P')_\nu - e^{(\lambda)* }_\nu (P+P')_\mu\right]\nonumber \\
&\qquad \quad -\tilde{B}(q^2)\left[ e^{(\lambda)*}_\mu q_\nu- e^{(\lambda)*}_\nu q_\mu\right] \nonumber\\
&\qquad \quad- 2  \tilde{C}(q^2) \frac{e^{(\lambda)*}\cdot q}{m_B^2-m_{f_2}^2} \left[ P_\mu q_\nu- P_\nu q_\mu\right],
\end{align}
where $ q_\alpha=P_\alpha' -P_\alpha$, $e^{(\lambda)*}_{ \alpha} =\frac{e^{(\lambda)*}_{ \alpha \beta} q^\beta}{m_B}$ and 
\begin{align*}
 \tilde{A}_3(q^2)&= \frac{m_B+m_{f_2}}{2 m_{f_2}} \tilde{A}_1(q^2)-\frac{m_B-m_{f_2}}{2 m_{f_2}} \tilde{A}_2(q^2).
\end{align*}
The tensor form factors can also be defined by the two following matrix elements
\begin{align*}
&\bra{f_2^\lambda(P)} \bar{u}(0) \sigma^{ \mu \nu} q_\nu b(0) \ket{B(P')}= \\
&-2i  \epsilon^{ \mu \nu \alpha \beta} P'_\nu P_\alpha e^{(\lambda)*}_\beta \tilde{T}_1(q^2), 
\end{align*}
\begin{align*}
&\bra{f_2^\lambda(P)} \bar{u}(0) \sigma^{ \mu \nu} \gamma_5 q_\nu b(0) \ket{B(P')} =\\
& \tilde{T}_2(q^2) \left[(m_B^2-m_{f_2}^2) e^{(\lambda)*\mu} -e^{(\lambda)*} \cdot P' (P'+P)^\mu\right] \\
&+ \tilde{T}_3(q^2) e^{(\lambda)*} \cdot P' \left[ q^\mu - \frac{q^2}{m_B^2-m_{f_2}^2} (P'+P)^\mu \right],
\end{align*}
which then leads to
\begin{align*}
\tilde{T}_1(q^2)&=\tilde{A}(q^2),\\
\tilde{T}_2(q^2)&=\tilde{A}(q^2)- \frac{q^2}{m_B^2-m_{f_2}^2}\tilde{B}(q^2),\\
\tilde{T}_3(q^2)&=\tilde{B}(q^2)+\tilde{C}(q^2).
\end{align*}
To get access to these form factors we use the two-point correlation function
\begin{align}
\Pi_{a}(q,P)&=i \int d^4x  e^{i q  x} \bra{f_2^\lambda(P)} T \{ \bar{q}_1(x)\Gamma_a b(x)j_B(0)\}\ket{0}, 
\label{correlation}
\end{align}
with the  Lorentz structures
\begin{align*}
\Gamma_\mu&=\gamma_\mu , \quad \Gamma_{\mu 5}=\gamma_\mu \gamma_5 ,\quad\Gamma_{\mu \nu 5}=\sigma_{\mu \nu} \gamma_5 .
\end{align*}
Here
\begin{align*}
j_B(0)=\bar{b}(0) i \gamma_5 q_2(0) 
\end{align*}
is the interpolating current for the B-meson. \\
The decay constant $f_B$ of the B-meson is defined by
\begin{align}
\bra{B(P')}\bar{b}(0) i \gamma_5 q_2(0) \ket{0}=\frac{f_B m_B^2}{m_b}.
\label{deco}
\end{align}
The standard procedure of light cone sum rules is to calculate the correlation function (\ref{correlation}) in two different ways. On the one hand, for large virtualities  we use operator product expansion (OPE) around the light cone 
so that we can represent the correlation function in terms of the light cone DAs, which are given in appendix \ref{das}. On the other hand we can insert a complete set of eigenstates with the quantum numbers of the B-meson and isolate the ground state.
These two different representations can be matched using dispersion relations and quark-hadron duality. Using Borel-transformation to eliminate subtraction terms and to suppress higher states leads to the final sum rules.\\
For the hadronic representation after inserting a complete set of eigenstates and isolating the ground state 
we get, e.g., for the vector current
\begin{align*}
\Pi_\mu(q,P)=&\frac{\bra{0} \bar{q}_1 \gamma_\mu b \ket{B} \bra{B} \bar{b} i \gamma_5 q_2 \ket{0}}{m_B^2-q^2} \\
&+ \sum_h \frac{\bra{0} \bar{q}_1 \gamma_\mu b \ket{h} \bra{h} \bar{b} i \gamma_5 q_2 \ket{0}}{m_h^2-q^2}.
\end{align*}
Inserting equations (\ref{fofa1}), (\ref{deco}) and rewriting the higher states into a dispersion integral over a spectral density, describing the excited and continuum states
we get
\begin{align*}
\Pi_\mu(q,P)=&\frac{2 f_B m_B^2  \epsilon_{\mu \nu \alpha \beta} e_{(\lambda)*}^{\nu} P'^\alpha P^\beta \tilde{V}(q^2)}{m_b(m_B+m_{f_2})(m_B^2-q^2)} \\
&+\int\limits_{s_0^h}^{\infty} \frac{ds}{s-q^2}\rho_\mu(s).
\end{align*}
Here $s_0^h$ is the threshold of the lowest continuum state.
Applying a Borel-transformation
\begin{align*}
\frac{1}{s-P'^2} \rightarrow e^{-s/M^2}, \qquad (P'^2)^n \rightarrow 0,
\end{align*}
we get for the vector case and the other two cases after the same procedure
\begin{widetext}
\begin{align*}
\Pi_\mu(q,P)&=2  f_B m_B^2 e^{-m_B^2/M^2}\frac{\tilde{V}(q^2)}{(m_B+m_{f_2})m_b} \epsilon_{\mu \nu \alpha\beta} \,e_{(\lambda)* }^\nu q^\alpha P^\beta ,\\
\Pi_{\mu 5}(q,P)&=\frac{i f_B m_B^2}{m_b} e^{-m_B^2/M^2}  \left[ (m_B+m_{f_2}) e^{(\lambda)*}_ \mu \tilde{A}_1(q^2)-(e^{(\lambda)*} \cdot q)(2P_\mu+q_\mu) \frac{\tilde{A}_2(q^2)}{m_B+m_{f_2}} \right. \\
&\left.  - 2 m_{f_2} \frac{e^{(\lambda)*}\cdot q}{q^2} q_\mu \left(\tilde{A}_3(q^2)-\tilde{A}_0(q^2)\right) \right], \\
\Pi_{\mu \nu 5}(q,P)&=\frac{ f_B m_B^2}{m_b} e^{-m_B^2/M^2} \left[ -\tilde{A}(q^2)\left((2P_\mu+q_\mu) e^{(\lambda)*}_\nu-(2P_\nu+q_\nu) e^{(\lambda)*}_\mu\right) \right. \\
&\left. - \tilde{B}(q^2)\left(e^{(\lambda)*}_\mu q_\nu-e^{(\lambda)*}_\nu q_\mu\right)-2 \tilde{C}(q^2) \frac{e^{(\lambda)*}\cdot q}{m_B^2-m_{f_2}^2}\left( P_\mu q_\nu-P_\nu q_\mu \right) \right],
\end{align*}
\end{widetext}
with $M^2$ being the Borel parameter.
For simplicity we do not write down the spectral densities.
Later we will use quark-hadron duality  to subtract these contributions from our OPE result. \\
For the OPE we contract the two $b$-quarks in (\ref{correlation}) using the quark propagator in a background field \cite{Belyaev:1994zk,Balitsky:1987bk}
\begin{align*}
&\bra{0} T\{ b^i(x) \bar{b}^j(0)  \} \ket{0}=-i \int \frac{d^4k}{(2\pi)^4} e^{-i k x} \frac{\slashed{k}+m_b}{m_b^2-k^2} \delta_{ij} \\
&-i g_s \int \frac{d^4k}{(2 \pi)^4} e^{- i k x} \int_0^1 dv\, G^{\mu \nu a}(vx) \left( \frac{\lambda^a}{2}\right)^{ij}  \\
&\left( \frac{\slashed{k}+m_b}{2(m_b^2-k^2)^2} \sigma_{\mu \nu} + \frac{1}{m_b^2-k^2} v x_\mu \gamma_\nu\right).
\end{align*}
So we get, e.g., for the vector current
\begin{align*}
\Pi_\mu(q,P)=&i \int \frac{d^4x d^4k}{(2\pi)^4} \frac{e^{ix(q-k)}}{m_b^2-k^2} \\
&\left( m_b \bra{f_2^\lambda(P)} \bar{q}_1(x) \gamma_\mu \gamma_5 q_2(0) \ket{0} \right.  \\
&\left.+ k^\nu \bra{f_2^\lambda(P)} \bar{q}_1(x) \gamma_\mu \gamma_\nu \gamma_5 q_2(0) \ket{0} \right. \\
&\left.+\int_0^1 dv \bra{f_2^\lambda(P)} \bar{q}_1(x) \gamma_\mu G^{\alpha \beta}(vx) \right.\\
&\left.\left( \frac{\slashed{k}+m_b}{2(m_b^2-k^2)} \sigma_{\alpha \beta}+\frac{v x_\alpha \gamma_\beta}{m_b^2-k^2}\right) \gamma_5 q_2(0) \ket{0}\right)
\end{align*}
After rewriting the Lorentz structures, if necessary,  the resulting matrix elements are expressed in terms of the light cone DAs from appendix \ref{das}.
After performing the $x$ and $k$ integration, the general structure, shown in simplified form looks like 
\begin{align}
\Pi_\mu(q,P)\sim\sum_{n=1}^5 \int\limits_0^1 du \frac{\mathcal{A}(u)}{D^n},
\label{qhdbsp}
\end{align}
where $\mathcal{A}(u)$ is one of the DAs from appendix \ref{das} and the denominator is
\begin{align*}
D=m_b^2-(q+uP)^2 .
\end{align*}
We have to write (\ref{qhdbsp}) as a dispersion integral in $P'^2$
\begin{align*}
\Pi_\mu(q,P)\sim\frac{1}{\pi} \int\limits_{m_b^2}^{\infty} \frac{ds}{s-P'^2} \text{Im}_s \mathcal{A}(s),
\end{align*}
which we can achieve by substituting
\begin{align*}
s(u)=&\frac{1}{u}(m_b^2-\bar{u}q^2+u \bar{u}m_{f_2}^2),\\ 
u(s)=&\frac{1}{2m_{f_2}^2}\bigg[m_{f_2}^2+q^2-s \\
&+\sqrt{(s-q^2-m_{f_2}^2)^2+4m_{f_2}^2(m_b^2-q^2)} \bigg],
\end{align*}
in the denominator (\ref{qhdbsp}), with $\bar{u}=1-u$ and perform a partial integration whenever the power of the denominator is larger than one.
Now the contributions of the excited and continuum states can be approximated using quark-hadron duality
\begin{align*}
\int\limits_{s_0^h}^{\infty}\frac{ds}{s-P'^2} \rho(s)\approx \frac{1}{\pi} \int\limits_{s_0}^{\infty} \frac{ds}{s-P'^2} \text{Im}_s \mathcal{A}(s),
\end{align*}
where $s_0$ is the duality threshold. 
For the final sum rules we use following shorthand notation for the DAs
\begin{align*}
\frac{d}{du} \hat{\mathcal{A}}(u)&=- \mathcal{A}(u),\\
\frac{d}{d \alpha_3} \tilde{\mathcal{T}}(\underline{\alpha})&=-\mathcal{T}(\underline{\alpha})
\end{align*}
with $\hat{\mathcal{A}}(0)=\hat{\mathcal{A}}(1)=\tilde{\mathcal{T}}(\alpha_3=0)=\tilde{\mathcal{T}}(\alpha_3=1)=0$. 
For $\hat{\hat{\mathcal{A}}}(u)$, $\tilde{\tilde{\mathcal{T}}}(\underline{\alpha})$ we have two derivatives etc. 
Performing a Borel-transformation  one obtains the final sum rules for the form factors
\begin{widetext}
\begin{align*}
\tilde{V}(q^2)=&\frac{(m_B+m_{f_2})m_b}{2 f_B m_B}  e^{m_B^2/M^2} \frac{1}{\pi} \int\limits_{m_b^2}^{s_0} ds \, e^{-s/M^2} \left[8(1-\delta_+) m_b m_{f_2}^2 f_{f_2} \text{Im}_s \hat{\hat{g}}_a(s)-2 m_{f_2} f_{f_2}^T \text{Im}_s \hat{A}(s) \right. \\
&\left. -(3m_b^2+1) m_{f_2}^3  f_{f_2}^T \text{Im}_s \hat{\mathbb{A}}(s)\right] , \\
\tilde{A}_1(q^2) =& \frac{m_b \, e^{m_B^2/M^2}}{f_B m_B (m_B+m_{f_2})} \frac{1}{\pi} \int\limits_{m_b^2}^{s_0} ds e^{-s/M^2} \left[ 2 m_b m_{f_2}^2 f_{f_2}  \text{Im}_s \hat{B}_1(s) - 4 m_{f_2}^3 m_b^2 f_{f_2}^T  \text{Im}_s \hat{\hat{C}}(s)-4 (1-\delta_+^T)m_{f_2}^3 f_{f_2}^T  \text{Im}_s \hat{\hat{h}}_\parallel^{(s)}(s) \right. \\
&\left. +8 m_bm_{f_2}^4 f_{f_2} \text{Im}_s \hat{\hat{\hat{C}}}_1(s)-2 m_{f_2}f_{f_2}^T (uP^2+Pq)\text{Im}_s \hat{A}(s)+(3m_b^2+1) m_{f_2}^3  f_{f_2}^T(uP^2+Pq) \text{Im}_s \hat{\mathbb{A}}(s) \right. \\
&\left. +16 m_{f_2}^3 f_{f_2}^T (uP^2+Pq) \text{Im}_s \hat{\hat{\hat{B}}}(s)  +8 \int\limits_0^ud \alpha_3 \int\limits_{\frac{u-\alpha_3}{1-\alpha_3}}^1 dv \,f_{f_2}^T m_{f_2}^5 \text{Im}_s \left(\tilde{\tilde{\mathcal{T}}}_1(\underline{\alpha})-\frac{m_{f_2}^2}{2} \tilde{\tilde{\mathcal{T}}}_2(\underline{\alpha})\right)\Bigg \vert_{\substack{\alpha_1=1-\alpha_2-\alpha_3 \\ \mathmakebox[2.25em][r]{\alpha_2=\frac{u-\alpha_3}{v}}}}\right] , \\
\tilde{A}_2(q^2)=&-\frac{(m_b+m_{f_2})m_b}{2f_B m_B} e^{m_B^2/M^2} \frac{1}{\pi} \int\limits_{m_b^2}^{s_0}ds \, e^{-s/M^2} \left[  8 m_b m_{f_2}^2 f_{f_2}\text{Im}_s \hat{\hat{A}}_1(s)+24 m_b^3 m_{f_2}^4 f_{f_2} \text{Im}_s \hat{\hat{\phi}}_4(s)+2 m_{f_2} f_{f_2}^T \text{Im}_s \hat{A}(s) \right. \\
&\left.-(3 m_b^2+1) m_{f_2}^3 f_{f_2}^T \text{Im}_s \hat{\mathbb{A}}(s)-8 (1-\delta_+^T)u m_{f_2}^3 f_{f_2}^T \text{Im}_s \hat{\hat{h}}_\parallel^{(s)}(s)+24 u m_b m_{f_2}^4 f_{f_2} \text{Im}_s \hat{\hat{\hat{C}}}_1(s) \right. \\
&\left.+4 u m_{f_2}^3 f_{f_2}^T\text{Im}_s \hat{\hat{C}}(s) - m_{f_2}^3 f_{f_2}^T (56+48(u Pq+q^2)) \text{Im}_s \hat{\hat{\hat{B}}}(s) \right.\\
&\left. +8 \int\limits_0^ud \alpha_3 \int\limits_{\frac{u-\alpha_3}{1-\alpha_3}}^1 dv \,f_{f_2}^T m_{f_2}^5 \left( 3 u \text{Im}_s \big(\tilde{\tilde{\mathcal{T}}}_1(\underline{\alpha})-\frac{m_{f_2}^2}{2} \tilde{\tilde{\mathcal{T}}}_2(\underline{\alpha})\big)-\frac{1}{m_{f_2}^2}  \text{Im}_s \big(\tilde{\tilde{\mathcal{T}}}_1(\underline{\alpha})-\frac{m_{f_2}^2}{2} \tilde{\tilde{\mathcal{T}}}_2(\underline{\alpha})\big)\right)\Bigg  \vert_{\substack{\alpha_1=1-\alpha_2-\alpha_3 \\ \mathmakebox[2.25em][r]{\alpha_2=\frac{u-\alpha_3}{v}}}}\right], \\
\tilde{A}_0(q^2)=&\frac{q^2}{2m_{f_2}} \left( \frac{m_b}{f_Bm_b}e^{m_B^2/M^2}\frac{1}{\pi} \int\limits_{m_b^2}^{s_0} ds \, e^{-s/M^2} \left[48 m_{f_2}^3 f_{f_2}^T (u P^2+Pq)\text{Im}_s \hat{\hat{\hat{B}}}(s)+24 m_b m_{f_2}^4 f_{f_2} \text{Im}_s \hat{\hat{\hat{C}}}(s)   \right. \right. \\
&\left. \left. -8 (1-\delta_+^T)  m_{f_2}^3 f_{f_2}^T \text{Im}_s\hat{\hat{h}}_\parallel^{(s)}(s)+4 m_{f_2}^3 f_{f_2}^T \text{Im}_s \hat{\hat{C}}(s)  \right. \right. \\
&\left. \left. +24 \int\limits_0^ud \alpha_3 \int\limits_{\frac{u-\alpha_3}{1-\alpha_3}}^1 dv \,  m_{f_2}^5 f_{f_2}^T \text{Im}_s \big( \tilde{\tilde{\mathcal{T}}}_1 (\underline{\alpha})-\frac{m_{f_2}^2}{2} \tilde{\tilde{\mathcal{T}}}_2(\underline{\alpha}) \big) \Big  \vert_{\substack{\alpha_1=1-\alpha_2-\alpha_3 \\ \mathmakebox[2.25em][r]{\alpha_2=\frac{u-\alpha_3}{v}}}} \right] + \frac{\tilde{A}_2(q^2)}{m_B+m_{f_2}}\right) \\
&+ \frac{m_b+m_{f_2}}{2 m_{f_2}} \tilde{A}_1(q^2)-\frac{m_b-m_{f_2}}{2 m_{f_2}} \tilde{A}_2(q^2) ,\\
\tilde{A}(q^2)=& -\frac{m_b }{2 f_B m_B} e^{m_B^2/M^2} \frac{1}{\pi} \int\limits_{m_b^2}^{s_0} ds \, e^{-s/M^2}\left[ -2 m_b m_{f_2} f_{f_2}^T \text{Im}_s \hat{A}(s)  +3m_b^3 m_{f_2}^3 f_{f_2}^T \text{Im}_s \hat{\mathbb{A}}(s)+16 m_b m_{f_2}^3 f_{f_2}^T \text{Im}_s \hat{\hat{\hat{B}}}(u) \right. \\
&\left.+ 8 (1-\delta_+) m_{f_2}^2 f_{f_2} \text{Im}_s \hat{\hat{g}}_a(s)-4 m_{f_2}^2 f_{f_2} \text{Im}_s \hat{\hat{A}}_1(s)+2 u m_{f_2}^2 f_{f_2} \text{Im}_s \hat{B}_1(s)+2(3m_b^2+1) m_{f_2}^4 f_{f_2} \text{Im}_s \hat{\hat{\phi}}_4(s) \right. \\
&\left.+ 2\int\limits_0^u d \alpha_3 \int\limits_{\frac{u-\alpha_3}{1-\alpha_3}}^1 dv \,m_{f_2}^2 f_{f_2} \left( \text{Im}_s \big( \mathcal{V}(\underline{\alpha})-\mathcal{A}(\underline{\alpha})\big)-4 m_{f_2}^2 \text{Im}_s \big( \tilde{\tilde{\mathcal{V}}}(\underline{\alpha})-\tilde{\tilde{\mathcal{A}}}(\underline{\alpha})  \big) \right. \right. \\
&\left. \left.+2u  m_{f_2}^2 \text{Im}_s\big( \tilde{\mathcal{V}}(\underline{\alpha})-\tilde{\mathcal{A}}(\underline{\alpha}) \big) \right) \Big  \vert_{\substack{\alpha_1=1-\frac{u-\alpha_3}{v}-\alpha_3 \\ \mathmakebox[1.0em][r]{\alpha_2=\frac{u-\alpha_3}{v}}}}\right],\\
\end{align*}
\begin{align*}
\tilde{B}(q^2)&= \frac{m_b}{f_B m_B}e^{m_B^2/M^2} \frac{1}{\pi} \int\limits_{m_b^2}^{s_0} ds \, e^{-s/M^2}\left[ -8 (1-\delta_+) m_{f_2}^2 f_{f_2} (uP^2+Pq)  \text{Im}_s \hat{\hat{g}}_a(s)+2 m_{f_2}^2 f_{f_2} \text{Im}_s \hat{B}_1(s) \right. \\
&\left.+4 \int\limits_0^u d \alpha_3 \int\limits_{\frac{u-\alpha_3}{1-\alpha_3}}^1 dv \, m_{f_2}^4 f_{f_2} \text{Im}_s\big( \tilde{\mathcal{V}}(\underline{\alpha})-\tilde{\mathcal{A}}(\underline{\alpha}) \big) \Big  \vert_{\substack{\alpha_1=1-\frac{u-\alpha_3}{v}-\alpha_3 \\ \mathmakebox[1.0em][r]{\alpha_2=\frac{u-\alpha_3}{v}}}}\right]+\tilde{A}(q^2) , \\
\tilde{C}(q^2)&=-\frac{(m_B^2-m_{f_2}^2)m_b}{2 f_Bm_B} e^{m_B^2/M^2} \frac{1}{\pi} \int \limits_{m_b^2}^{s_0} ds \, e^{-s/M^2} \left[ -48 m_b m_{f_2}^3 f_{f_2}^T \text{Im}_s \hat{\hat{\hat{B}}}(s)+8 (1-\delta_+) m_{f_2}^2 f_{f_2} \text{Im}_s \hat{\hat{g}}_a(s) \right. \\
&\left. +8 m_{f_2}^2 f_{f_2} \text{Im}_s \hat{\hat{A}}_1(s) - 6(4 m_b^2+1)f_{f_2} \text{Im}_s \hat{\hat{\phi}}_4(s)+3\int\limits_0^u d \alpha_3 \int\limits_{\frac{u-\alpha_3}{1-\alpha_3}}^1 dv \, m_{f_2}^2 f_{f_2} \text{Im}_s\big( \tilde{\tilde{\mathcal{V}}}(\underline{\alpha})-\tilde{\tilde{\mathcal{A}}}(\underline{\alpha}) \big) \Big  \vert_{\substack{\alpha_1=1-\alpha_2-\alpha_3 \\ \mathmakebox[2.25em][r]{\alpha_2=\frac{u-\alpha_3}{v}}}}\right].
\end{align*}
\end{widetext}

\section{Numerical Analysis and discussion}

\begin{figure*}
 \begin{minipage}{0.45\textwidth}
  \includegraphics[width=\textwidth]{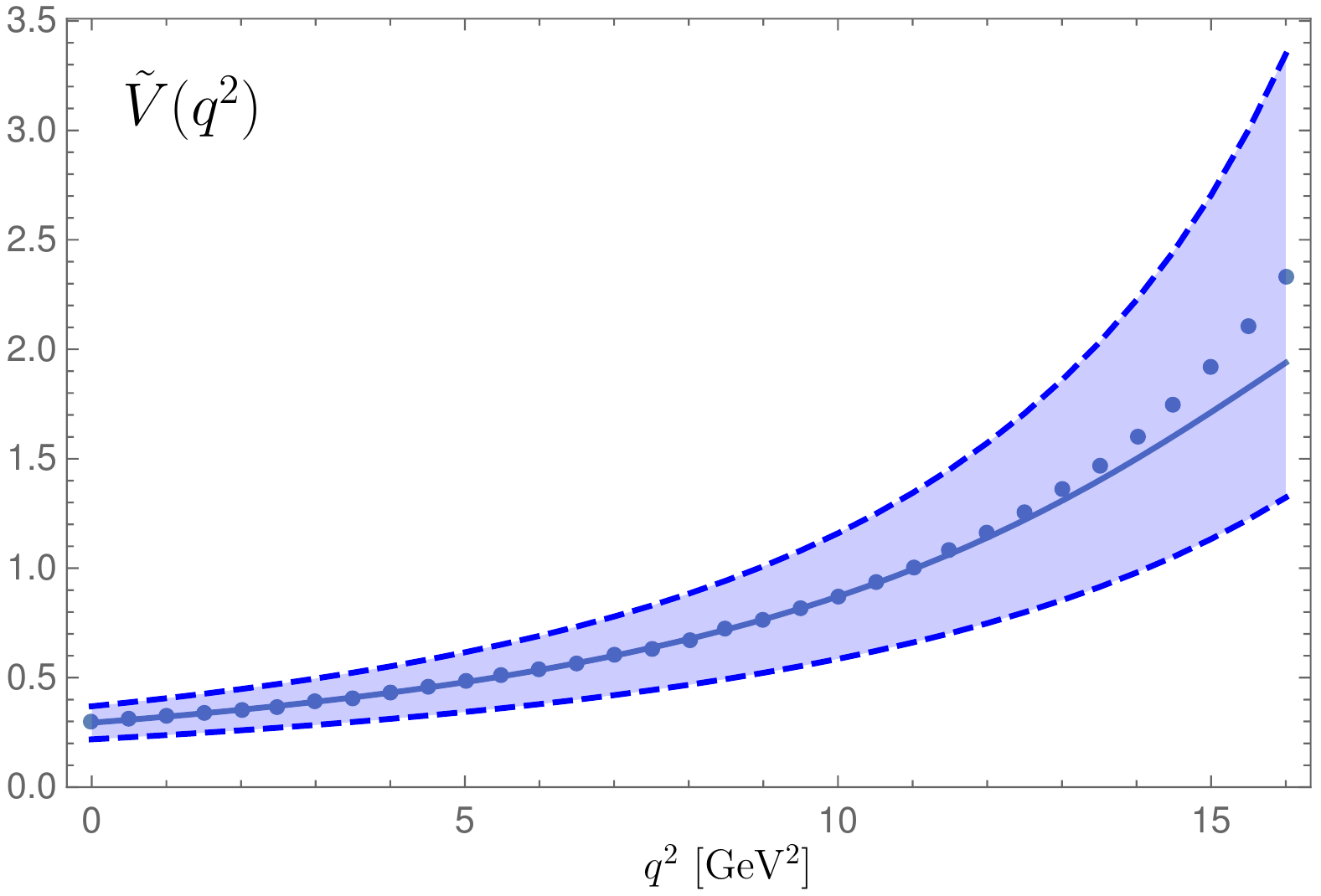}
 \end{minipage}
 \begin{minipage}{0.45 \textwidth}
  \includegraphics[width=\textwidth]{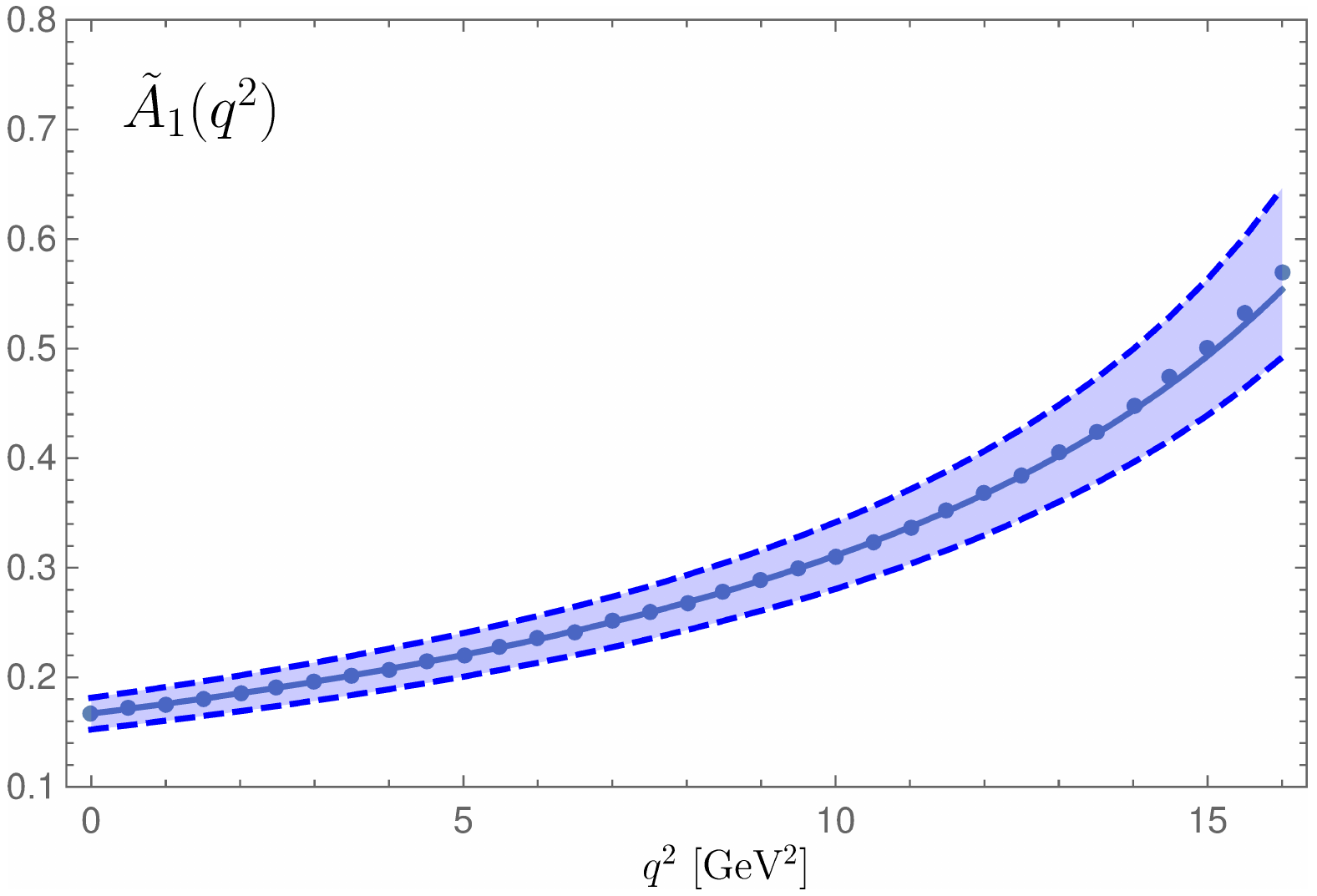}
 \end{minipage}
 \begin{minipage}{0.45\textwidth}
  \includegraphics[width=\textwidth]{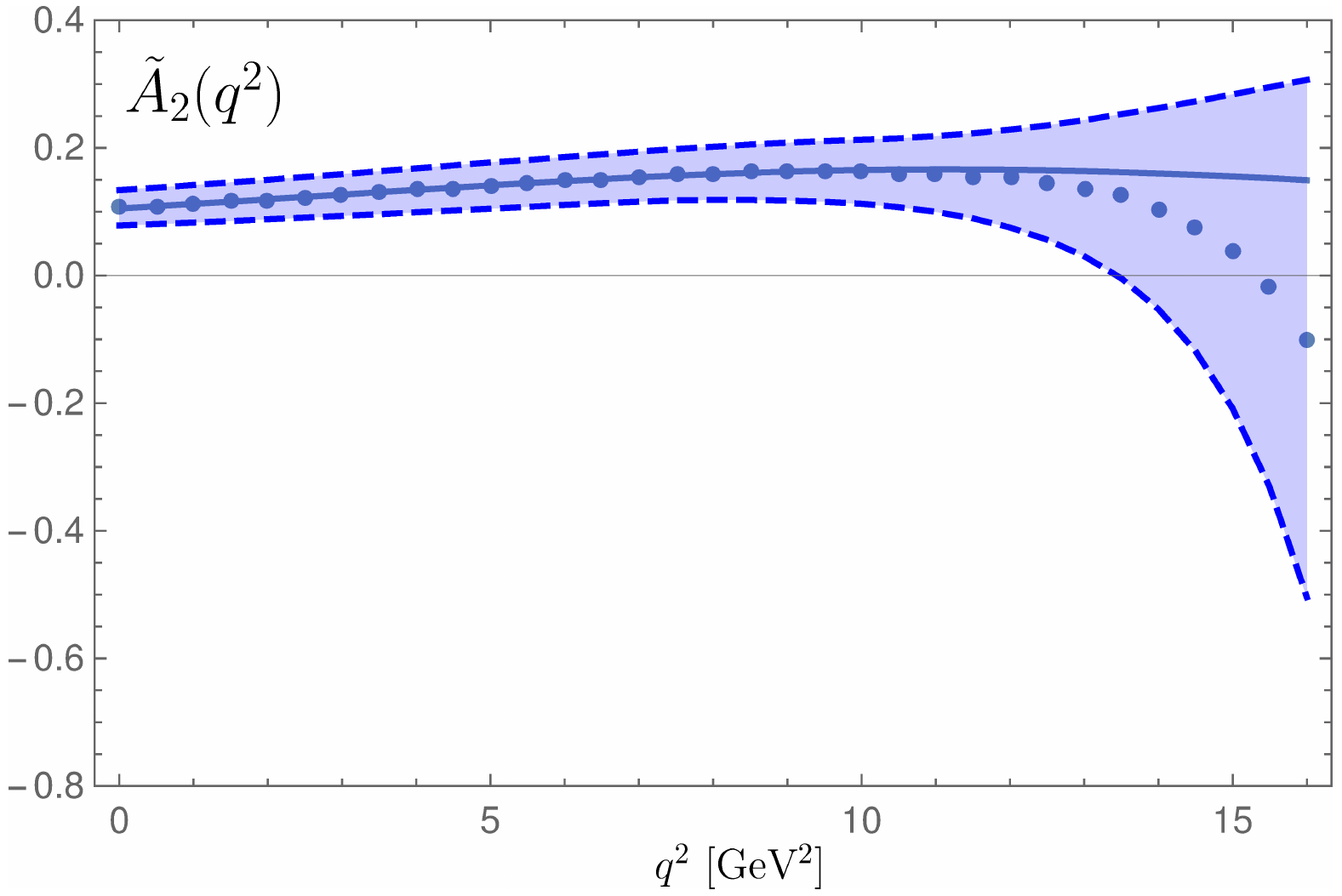}
 \end{minipage}
 \begin{minipage}{0.45 \textwidth}
  \includegraphics[width=\textwidth]{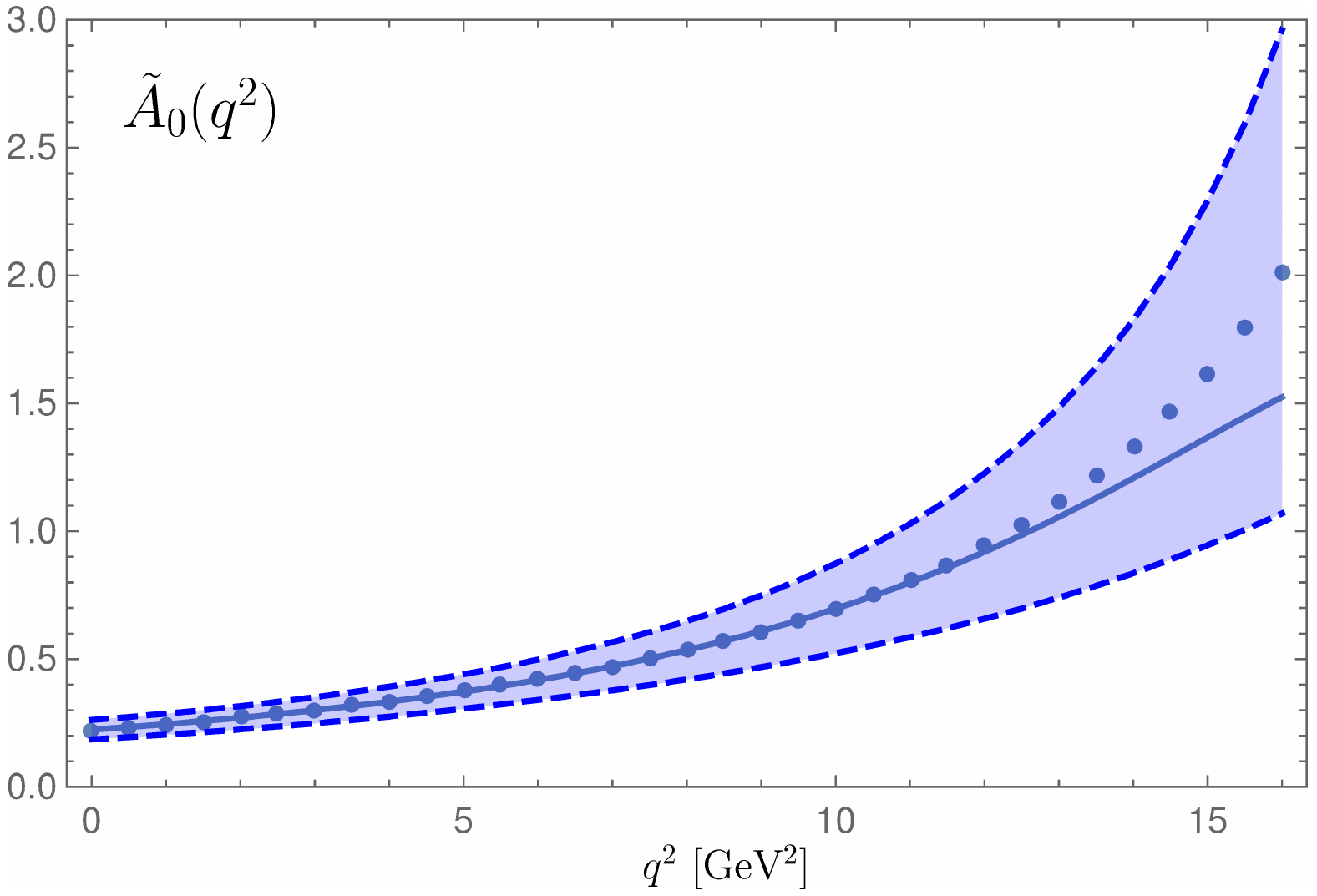}
 \end{minipage}
 \begin{minipage}{0.45\textwidth}
  \includegraphics[width=\textwidth]{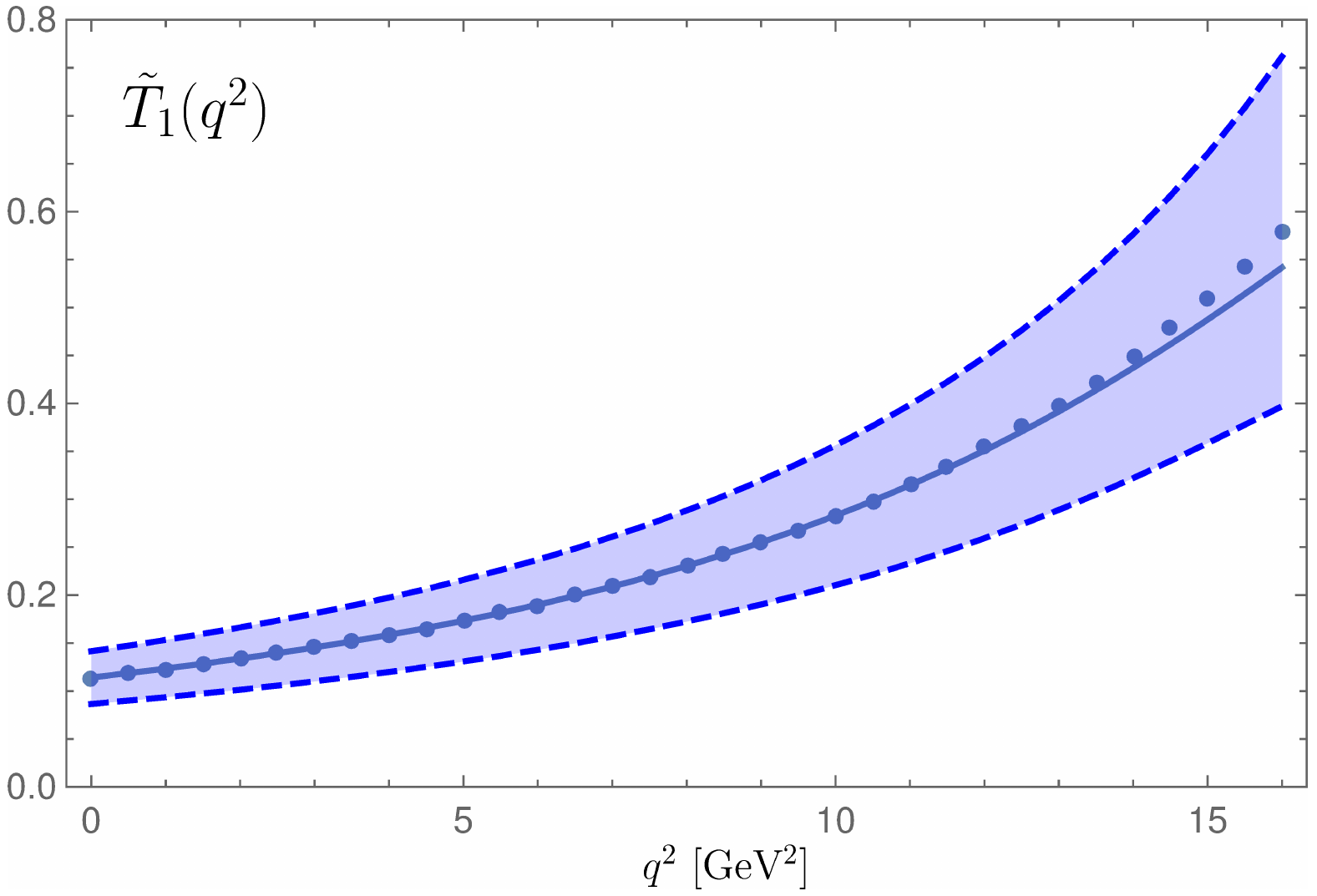}
 \end{minipage}
 \begin{minipage}{0.45 \textwidth}
  \includegraphics[width=\textwidth]{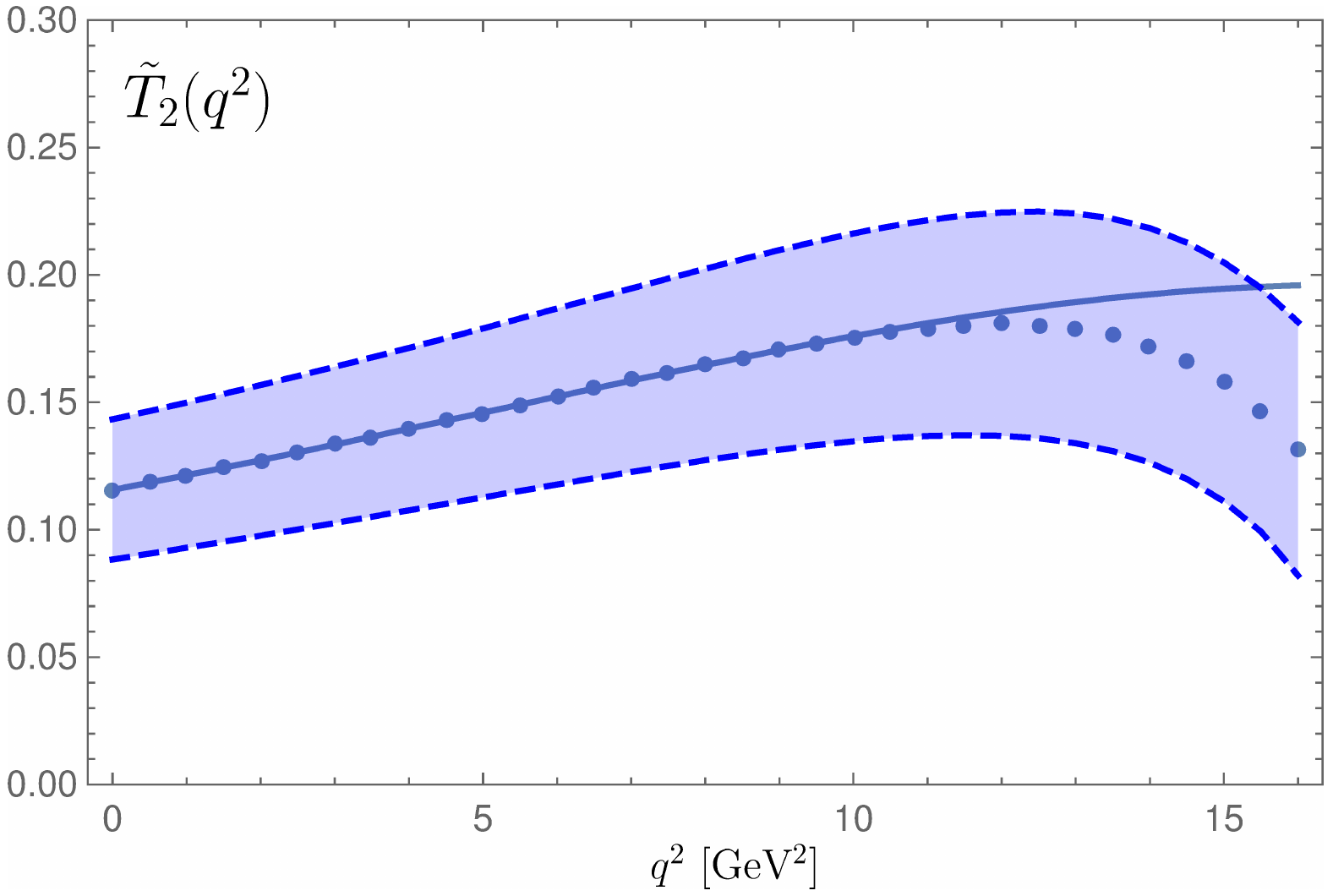}
 \end{minipage}
 \caption{$q^2$ dependence of the form factors. The solid line gives the central value of the fit to the sum rule results (dots) and the dashed lines are the uncertainties from varying the input parameters.}
 \label{qdep}
\end{figure*}

For the numerical analysis we use the following input values for the masses \cite{1674-1137-40-10-100001}
\begin{align*}
 m_{f_2}&=1.275\text{GeV} ,\qquad m_B=5.279 \text{GeV} ,
\end{align*}
and for the decay constants at a scale of $\mu=1$ GeV we use \cite{1007.3541v1,Braun2016} 
\begin{align*}
 f_{f_2}=0.101(10) \text{GeV}, \qquad  f_{f_2}^T=0.117(25) \text{GeV}  .
\end{align*}
We use the pole b-quark mass, as always for LCSR, given by $m_b=4.8(1)$ GeV and for the B-meson decay constant $f_B$ we use the tree level sum rule from \cite{Dominguez:1987ea} .
All the scale dependent parameters are evaluated at the factorization scale $\mu_f=\sqrt{m_B^2-m_b^2}$. We choose the Borel parameter window to be $M^2=4-8 \text{ GeV}^2$ and  the duality threshold  $s_0=35.5 \pm 2 \text{ GeV}^2$, which 
is consistent with other studies of the B-meson \cite{hep-ph/9805422v2}. All the other input values are given in appendix \ref{das}.

\begin{figure*}[ht]
\scalebox{0.93}{
\begin{picture}(800,240)(0,-5)

\put(-5,0){\epsfig{file=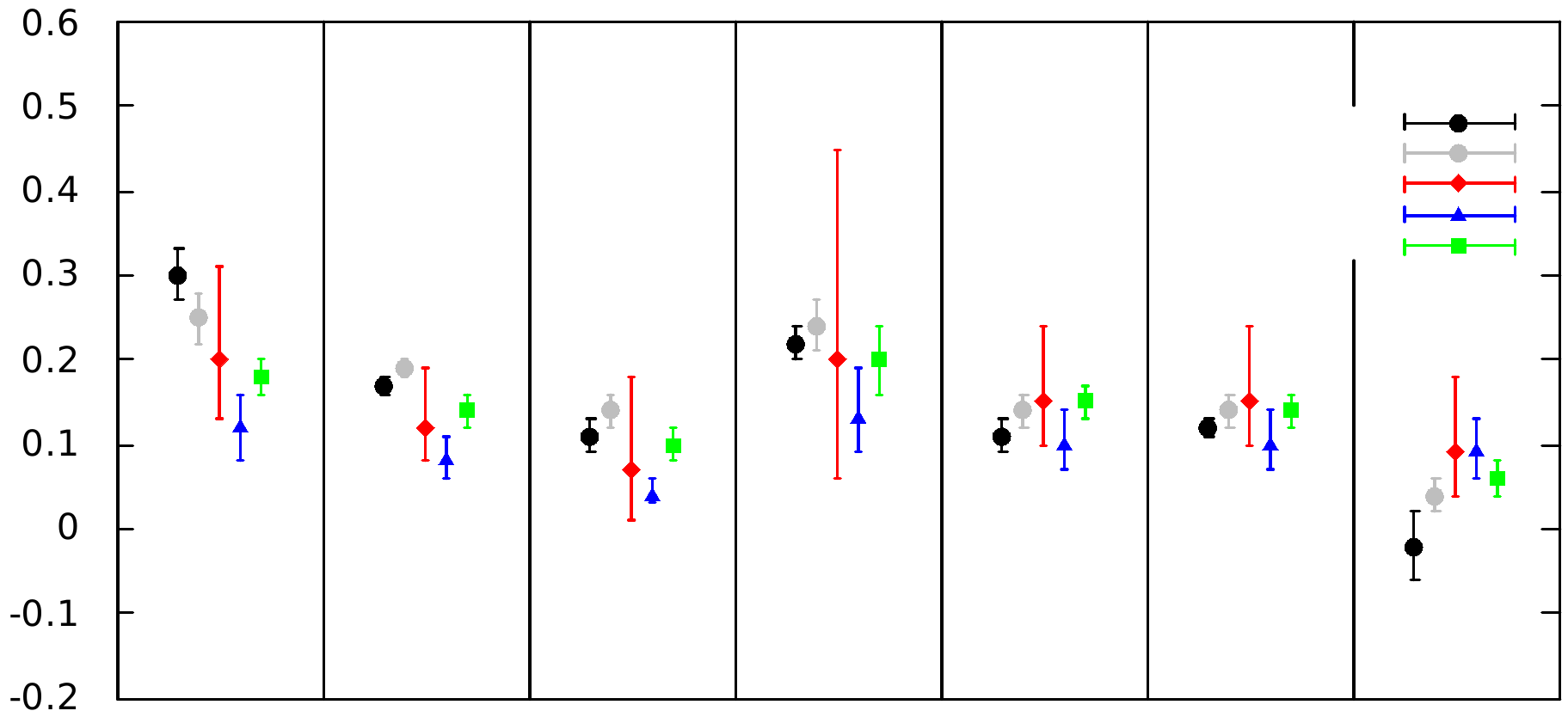,scale=1}}

\put(50,-10){ { \large$\tilde{V}(0)$}}
\put(125,-10){\large $\tilde{A}_1(0)$}
\put(195,-10){\large $\tilde{A}_2(0)$}
\put(265,-10){\large $\tilde{A}_0(0)$}
\put(335,-10){\large $\tilde{T}_1(0)$}
\put(405,-10){\large $\tilde{T}_2(0)$}
\put(475,-10){\large $\tilde{T}_3(0)$}
\put(430,166){Ref. \cite{1008.5326v2}}
\put(430,156){Ref. \cite{1010.2944v3}}
\put(430,176){Ref. \cite{Wang:2010tz}}
\put(382,187.5){Without mass terms}
\put(395.5,197.5){With mass terms}
\put(490,-20){}
\end{picture}
}
   \caption{The values of the form factors for $q^2=0$ from different theoretical approaches.}
   \label{figure1}
\end{figure*}
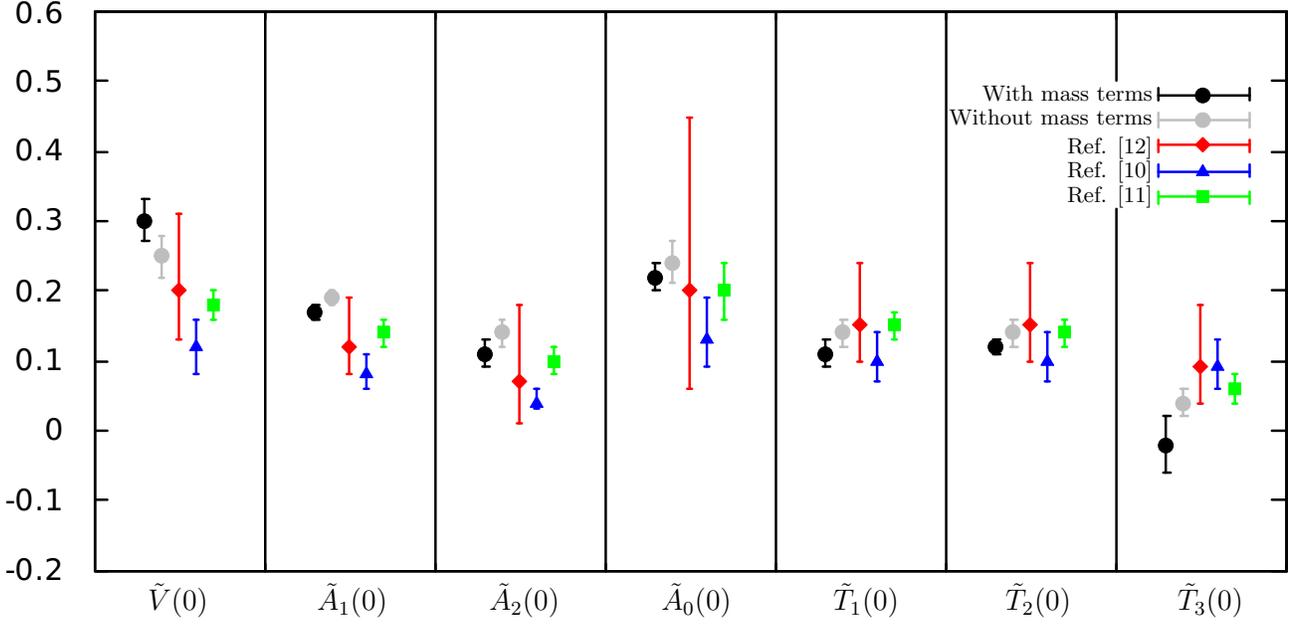

The LCSR are assumed to give a reasonable approximation up to $q^2 
\leq  q_{\text{max}}^2=(m_B-m_{f_2})^2= 16.07 $ GeV$^2$.
To avoid fitting artefacts, we limit the actual fit range to $0\leq q^2 
\leq10$ GeV$^2$. The deviations from the fit curves for large $q^2$ 
in fact indicate the break down of the approximation.
We choose a parameterization for the form factors
with the three parameters $\tilde{F}(0)$, $a$ and $b$
\begin{align}
F(q^2)=\frac{\tilde{F}(0)}{1-a (q^2/m_B^2)+b(q^2/m_B^2)^2}.
\label{fitform}
\end{align}
\begin{table}[ht]
\renewcommand{\arraystretch}{1.2}
\begin{tabular}{ |c| c| c |c| }
\hline
  Form Factor & $\tilde{F}(0)$ & a & b\\
  \hline
  $\tilde{V} $ & $0.30\pm 0.03 $ & $2.38 \pm0.4 $ &$1.50 \pm0.73$\\
  $\tilde{A}_1 $& $0.17\pm0.01 $&$1.41\pm 0.50$ &$0.35 \pm1.40$\\
  $\tilde{A}_2 $& $0.11\pm0.02 $&$1.84 \pm 1.46 $& $2.30\pm4.09$\\
  $\tilde{A}_0 $& $0.22\pm0.02 $& $2.57 \pm0.77$&$1.89 \pm 2.23$\\
  $\tilde{T}_1 $& $0.11\pm0.02 $& $2.14\pm 1.14$ &$1.34\pm3.19$\\
  $\tilde{T}_2$& $0.12\pm0.01 $& $1.35 \pm 1.24$& $1.11 \pm3.39$\\
  $\tilde{T}_3 $& $-0.02\pm0.04 $&$1.94 \pm 17.51$ & $0.71 \pm 49.40$\\
\hline
  \end{tabular}
\caption{Results from fitting the $B \rightarrow f_2(1270)$ form factors obtained by LCSR to the three parameter form in (\ref{fitform}).
}
\label{table1}
\end{table}We perform a weighted fit using as weights the uncertainties from varying 
the input parameters and add the errors in quadrature.
The cited errors indicate an increase of $\chi^2$ by 1. 
For asymmetric errors we take the mean value and shift the central value by the 
difference of the asymmetric error and the mean value to get
symmetric errors. As one can see from figure 
\ref{qdep} our $\chi^2/$ d.o.f. is nearly zero for all
form factors and $q^2 \leq 10$ GeV$^2$, indicating that the parameterization 
(\ref{fitform}) is a very efficient one.
We do not show any $q^2$ dependence of the form factor $\tilde{T}_3(q^2)$ 
because this form factor is close to zero in the whole fitting range due 
to the fact that $\tilde{B}(q^2)$ and $\tilde{C}(q^2)$ have nearly the same magnitude but different signs.
Our results can be found in table \ref{table1} and in figures \ref{qdep}, \ref{figure1}.
We observe that the contributions from the mass terms $\mathbb{A}(u)$ and $\phi_4(u)$ to the form factors are not negligible as can already be seen in figure \ref{figure1}.   
More precisely the effect of these mass terms is for all the form factors less than $30\%$  for $q^2=0$. For $q^2\neq0$ the contributions of the meson mass terms to the  form factors $\tilde{V}(q^2)$,  $\tilde{A}_1(q^2)$ and $\tilde{A}_0(q^2)$ stays under $30 \%$ . 
For the form factors $\tilde{A}_2(q^2)$, $ \tilde{T}_1(q^2)$, $\tilde{T}_2(q^2)$ and $\tilde{T}_3(q^2)$ the effect of the meson mass terms 
increases for higher values of $q^2$.
Worth mentioning is the form factor $\tilde{A}_0(q^2)$, which depends on $\tilde{A}_1(q^2)$ and $\tilde{A}_2(q^2)$ but, due to cancellations the effect of the meson mass terms is less than $13 \%$ for the whole range  of $q^2$. 
 The comparison with other theoretical approaches, which is illustrated in figure \ref{figure1}
by the $q^2=0$ values of the form factors, illustrates the improved precision we achieved. 
This comparison requires, however, some explanations. The method used in \cite{1008.5326v2} is a calculation within a specific ``pQCD'' approach based on $k_{\perp}$ factorization.
The discrepancies between their results and ours (black bullets) is substantially larger than 
the systematic uncertainty we expect for our LCSR calculation. Therefore, we conclude that we disagree 
with the findings of \cite{1008.5326v2}. In contrast \cite{1010.2944v3,Wang:2010tz} are also LCSR calculations 
which allows to trace back the discrepancies to the fact that we have calculated higher contributions. 
In all cases the error bars represent the variation observed when the LCSR input parameters are varied in 
reasonable bounds. They do not include any estimate of neglected higher order terms. Therefore,
\cite{1010.2944v3} should be compared to our grey bullets which do not contain meson mass corrections as these where 
also not taken into account in \cite{1010.2944v3}. The difference between our grey bullets and 
the green squares shows that the higher twist contributions and three particle DAs we take into account  
make a significant difference, especially for $\tilde V(0)$, though not a very large one. The same can be 
said of the meson mass terms, comparing our grey and black bullets. Thus, one can conclude that to reach high precision
all these effects have to be included and that our results are in fact far more precise than earlier 
calculations even though this does not show up in all cases in the cited error bars.  

To summarize, we calculated the $B \rightarrow f_2(1270)$ form factors with LCSR using chiral even and chiral odd tensor meson DAs, including for the first time twist four meson mass terms. 
We observe that these mass terms have a noticeable impact on the sum rules and should be taken into account in future studies. 
Especially for the region 
of $q^2 \neq 0$ these mass terms can play an important role. 
The effects of still higher twist terms are probably smaller then the uncertainties 
arising from the choice of the Borel parameter, which is illustrated by the cited 
error bars. However, this can only be checked by future calculations. 
In such future investigations we would, e.g., also consider additional SU(3) breaking terms. 
Especially for decays involving a strange quark such SU(3) breaking terms can 
probably yield important contributions.

\section{Acknowledgments}
We appreciate helpful discussions with V. M. Braun. This work was supported by Deutsche Forschungsgemeinschaft (DFG) with the grant SFB/TRR 55.
\appendix
\section{Distribution amplitudes} \label{das}
In previous studies the chiral even quark-antiquark light-cone DAs for the $f_2$-meson were defined as matrix elements of nonlocal light-ray operators \cite{hep-ph/0012220v1,1007.3541v1,Braun2016}
\begin{align}
&\bra{f_2(P, \lambda)} \bar{q}(z_2n) \gamma_\mu q(z_1 n)\ket{0} = \nonumber \\
&f_{f_2} m^2_{f_2} \left[ \frac{e^{(\lambda)*}_{nn}}{(pn)^2} p_\mu \int_0^1 du  \, e^{i z_{12}(pn)} \phi_2(u,\mu) \right.\nonumber \\
&\left.+ \frac{e_{\perp \mu n}^{ (\lambda)*}}{pn} \int_0^1 du  \, e^{i z_{12} (pn)} g_v(u,\mu) \right. \nonumber \\
& \left. - \frac{1}{2} n_\mu m_{f_2}^2 \frac{e_{nn}^{(\lambda)*}}{(pn)^3}  \int_0^1 du  \,e^{i z_{12} (pn)} g_4(u,\mu) \right] , \label{even1} \\
&\bra{f_2(P,\lambda)} \bar{q}(z_2n) \gamma_\mu \gamma_5 q(z_1n)\ket{0}=\nonumber \\
&-i f_{f_2} m_{f_2}^2 \left(1-\delta_+ \right) \epsilon_{\mu \nu \alpha \beta} \frac{n^\nu p^\alpha}{pn} \frac{e^{(\lambda)* }_{\beta n}}{pn} \nonumber \\
&\int_0^1 du \, e^{i z_{12} (pn)} g_a(u,\mu). \label{even2} 
\end{align}
In the same manner we can define the chiral odd DAs  \footnote{In Ref. \cite{1007.3541v1} they already defined the chiral odd DAs but without the mass terms and SU(3) breaking terms.  }
\begin{align}
&\bra{f_2(P,\lambda)} \bar{q}(z_2n)  q(z_1n) \ket{0}= \nonumber \\
&f^T_{f_2} \frac{e_{n n}^{(\lambda)*}}{(pn)^2} m_{f_2}^3 \left(1-\delta_+^T \right) \int_0^1 du\, e^{i z_{12}(pn)} h_{\parallel}^{(s)}(u,\mu) ,\label{odd2} 
\end{align}
\begin{align}
&\bra{f_2(P,\lambda)} \bar{q}(z_2n)  \sigma_{\mu \nu} q(z_1n) \ket{0}= \nonumber\\
&i f^T_{f_2} \left[ m_{f_2} \frac{(e_{\perp n \mu}^{ (\lambda)*}p_\nu-e^{ (\lambda)*}_{\perp n \nu}p_\mu)}{pn} \int_0^1 du \, e^{i z_{12}(pn)}\phi_{\perp} (u,\mu) \right. \nonumber \\
&\left. +m_{f_2}^3 (p_\mu n_\nu -p_\nu n_\mu) \frac{e_{n n}^{(\lambda)*}}{(pn)^3} \int_0^1du \, e^{i z_{12} (pn)} h_{\parallel}^{(t)}(u,\mu)  \right.  \nonumber\\
&\left. +\frac{1}{2} (e_{\perp n \mu}^{ (\lambda)*} n_\nu -e_{\perp n\nu}^{ (\lambda)*} n_\mu) \frac{m_{f_2}^3}{(pn)^2} \int_0^1du \, e^{i z_{12} (pn)} h_4(u,\mu) \right] , \label{odd1} 
\end{align}
with $e^{(\lambda)*}_{nn}=e_{\alpha \beta}^{(\lambda)*} n^\alpha n^\beta$ and we use the shorthand notation $z_{12}=\bar{u}z_1+u z_2$.
The polarization tensor $e_{\alpha \beta}^{(\lambda)}$ is traceless, symmetric and satisfies the condition $e_{\alpha \beta}^{(\lambda)}P^\alpha=0$.
Further we have
\begin{align*}
e^{ (\lambda)*}_{\perp \mu n}&\equiv g_{\mu \nu}^{\perp} e_{\nu n}^{(\lambda)*} =e_{\mu n}^{(\lambda)*}-\frac{e_{nn}^{(\lambda)*}}{(pn)}p_\mu +\frac{1}{2} n_\mu e_{n n}^{(\lambda)*} \frac{m_{f_2}^2}{(pn)^2},\\
g_{\mu \nu}^\perp&=g_{\mu \nu} - \frac{1}{pn} \left(n_\mu p_\nu+n_\nu p_\mu \right),
\end{align*}
where the vectors $n_\mu$ and $p_\mu=P_\mu - \frac{n_\mu}{2}\frac{m_{f_2}^2}{pn}$ are light-like, $n^2=p^2=0$. 
The SU(3) breaking terms are parametrized by
\begin{align*}
\delta_\pm= \frac{f_{f_2}^T}{f_{f_2}} \frac{m_{\bar{q}} \pm m_q}{m_{f_2}}, \qquad \delta_\pm^T=\frac{f_{f_2}}{f^T_q} \frac{m_{\bar{q}} \pm m_q}{m_{f_2}}.
\end{align*}

Close to the light cone $x^2 \rightarrow 0$ the operator product expansion (OPE) of the chiral odd DAs takes the form
\begin{align*}
 &\bra{f_2(P, \lambda)} \bar{q}(x) \sigma_{\mu \nu} q(-x) \ket{0}=\\
 &  i f^T_q \left[ m_{f_2}\frac{(e^{(\lambda)*}_{x\mu} P_\nu-e_{x \nu}^{(\lambda)*}P_\mu)}{(Px)} \int_0^1 du \, e^{i \xi (Px)}  \left[ A(u) \right. \right. \\
& \left. \left. + \frac{1}{4} x^2 m_{f_2}^2 \mathbb{A}(u) \right] \right. \\
& \left.+ m_{f_2}^3 (P_\mu x_\nu - P_\nu x_\mu) \frac{e_{xx}^{(\lambda)*}}{(Px)^3} \int_0^1 du  \,e^{i \xi (Px)} B(u) \right. \\
&\bra{f_2(P,\lambda)}\bar{q}(x)q(-x) \ket{0}=\\
&f^T_q \frac{e_{xx}^{(\lambda)*}}{(Px)^2} m_{f_2}^3 \left(1-\delta_+^T \right) \int_0^1 du \, e^{i \xi (Px)} h_\parallel^{(s)}(u),
\end{align*}
with the new two-particle twist-four DA $\mathbb{A}(u)$ that can be expressed in terms of the other DAs using QCD EOM, see below and 
$\xi=2u-1$.
By comparing to equations (\ref{odd1}) and (\ref{odd2}) we find
\begin{align*}
A(u)&= \phi_\perp(u), \\
B(u)&=h_\parallel^{(t)}(u)-\frac{\phi_\perp(u)}{2}-\frac{h_4(u)}{2},\\
C(u)&=h_4(u)-\phi_\perp(u).
\end{align*}
The OPE for the chiral even DAs can be found in \cite{Braun2016}.\\
We take the three-particle quark-antiquark-gluon DAs from Ref. \cite{Braun2016}
\begin{widetext}
 \begin{align*}
&\bra{f_2(P,\lambda)} \bar{q}(z_3n)i gG_{\mu \nu}(z_2 n) \gamma_\alpha q(z_1 n) \ket{0}=  -f_{f_2} m_{f_2}^2 \frac{p_\alpha}{pn} \left[ p_\mu e_{\perp n \nu}^{(\lambda)}- p_\nu e_{\perp n \mu}^{(\lambda)} \right]\int \mathcal{D} \alpha \, e^{ i pn \sum \alpha_k z_k} \mathcal{V}(\underline{\alpha}) +  \cdots ,\\
&\bra{f_2(P,\lambda)} \bar{q}(z_3n) g \widetilde{G}_{\mu \nu}(z_2 n) \gamma_\alpha \gamma_5 q(z_1 n) \ket{0}=  - f_{f_2} m_{f_2}^2 \frac{p_\alpha}{pn} \left[ p_\mu e_{\perp n \nu}^{(\lambda)}- p_\nu e_{\perp n \mu}^{(\lambda)} \right] \int \mathcal{D} \alpha \, e^{ i pn \sum \alpha_k z_k} \mathcal{A}(\underline{\alpha})+ \cdots ,\\
\intertext{ and we define a new one for tensor structures}
&\bra{f_2(P,\lambda)} \bar{q}(z_3 n) \sigma_{\alpha\beta} g  G_{\mu \nu}(z_2 n)q(z_1 n) \ket{0}= \\
& -f_{f_2}^{T} \frac{e_{nn}^{(\lambda)*}}{2 (pn)^2}m_{f_2}^{3} \left[ p_\alpha p_\mu g_{\beta \nu}^{\perp}-p_\beta p_\mu g_{\alpha \nu}^\perp-p_{\alpha} p_\nu g_{\beta \mu}^\perp+ p_\beta p_\nu g_{\alpha \mu}^\perp \right] \int \mathcal{D} \alpha \, e^{ i pn \sum \alpha_k z_k} \mathcal{T}_1(\underline{\alpha}) \\
&+f_{f_2}^{T} \frac{m_{f_2}^{3}}{2} \left[p_\alpha p_\mu e_{ \nu \beta}^{\perp(\lambda)*}-p_\beta p_\mu e_{ \nu \alpha}^{\perp(\lambda)*}  - p_\alpha p_\nu e_{ \mu \beta}^{\perp(\lambda)*}  +p_\beta p_\nu e_{ \mu \alpha}^{\perp(\lambda)*}  \right] \int \mathcal{D} \alpha \, e^{ i pn \sum \alpha_k z_k} \mathcal{T}_2(\underline{\alpha})+ \cdots ,
\end{align*}
\end{widetext}
with $ e_{ \mu \nu}^{\perp(\lambda)*}= e_{ \mu' \nu'}^{(\lambda)*} g^\perp_{\mu' \mu} g^\perp_{\nu' \nu}$.
For the asymptotic form of the three-particle DAs we take \cite{hep-ph/9802299v2,hep-ph/0306057v1}
\begin{align*}
\mathcal{V}(\alpha)&=360 \alpha_1\alpha_2^2 \alpha_3 \left[\xi_3+\frac{1}{2} \omega_3 (7 \alpha_2-3)\right], 
\end{align*}
\begin{align*}
\mathcal{A}(\alpha)&=360 \alpha_1\alpha_2^2 \alpha_3 \left[\frac{1}{2} \widetilde{\omega}_3 (\alpha_1-\alpha_3)\right], \\
\mathcal{T}_{1/2}(\alpha)&=360 \alpha_1\alpha_2^2 \alpha_3 \left[\xi_3^{\mathcal{T}_{1/2}}+\frac{1}{2} \omega_3^{\mathcal{T}_{1/2}} (7 \alpha_2-3)\right] .
\end{align*}
The constants $\xi_3, \text{ } \omega_3$ and $\widetilde{\omega}_3$ have been determined in \cite{Braun2016} by using QCD sum rules and are at a scale of 1 GeV
\begin{align*}
 \xi_3=0.15(8), \qquad \omega_3=-0.2(3), \qquad \widetilde{\omega}_3=0.06(1).
 \end{align*}
Using QCD sum rules we get 
\begin{align*}
\left( \frac{m_{f_2}^2 \xi_3^{\mathcal{T}_2}}{2}-\xi_3^{\mathcal{T}_1}\right)&=0.16(3), \\
\left( \frac{m_{f_2}^2 \omega_3^{\mathcal{T}_2}}{2}-\omega_3^{\mathcal{T}_1}\right)&=-0.33(16). \\
\end{align*}
In equations (\ref{even1})-(\ref{odd2}), the two particle DAs $\phi_2(u)$, $\phi_\perp(u)$ are leading twist two, $g_v(u), \text{ } g_a(u),\text{ } h^{(t)}_\parallel(u), \text{ }h^{(s)}_\parallel(u)$ are collinear twist three and $g_4(u),\text{ }h_4(u)$ are twist four.\\
By using the EOM \cite{hep-ph/9802299v2} we can represent the twist three DAs, including the SU(3) breaking terms in terms of the leading DAs and three-particle DAs
\begin{align*}
&(1-\delta_+)g_a(u)=\int_0^u dv \frac{\Omega(v)}{\bar{v}}-\int_u^1 dv \frac{\Omega(v)}{v} ,\\
&g_v(u)=\int_0^u dv \frac{\Omega(v)}{\bar{v}}+\int_u^1 dv \frac{\Omega(v)}{v}+\delta_+ \phi_\perp(u) \nonumber\\
&- \frac{d}{du} \int_0^u d\alpha_1 \int_0^{\bar{u}}d \alpha_3 \frac{1}{\alpha_2} \mathcal{V}(\underline{\alpha}) \nonumber \\
&- \int_0^u d\alpha_1 \int_0^{\bar{u}} d\alpha_3 \frac{1}{\alpha_2}\left( \frac{d}{d \alpha_1}+\frac{d}{d\alpha_3}\right) \mathcal{A}(\underline{\alpha}),
\end{align*}
with 
\begin{align*}
&\Omega(u)=\phi_2(u)+(\delta_- +\delta_+ (2u-1)) \phi_\perp'(u) \\
&- \frac{1}{2} \frac{d}{du} \int_0^u d\alpha_1 \int_0^{\bar{u}}d \alpha_3 \frac{1}{\alpha_2}\left( \alpha_1 \frac{d}{d\alpha_1} + \alpha_3 \frac{d}{d\alpha_3}\right) \mathcal{V}(\underline{\alpha}) \\
&- \frac{1}{2} \frac{d}{du} \int_0^u d\alpha_1 \int_0^{\bar{u}}d \alpha_3 \frac{1}{\alpha_2}\left(\alpha_1 \frac{d}{d \alpha_1} - \alpha_3 \frac{d}{d \alpha_3} \right) \mathcal{A}(\underline{\alpha}),
\end{align*}
and
\begin{align*}
&(1-\delta_+^T)h_\parallel^{(s)}(u)=\frac{1}{2}\left( \int_0^u dv \, \frac{\Phi(v)}{\bar{v}} - \int_u^1 dv \, \frac{\Phi(v)}{v}  \right) ,\\
&h_\parallel^{(t)}=\frac{1}{2} (u- \bar{u}) \left(\int_0^u dv \, \frac{\Phi(v)}{\bar{v}} -\int_u^1 dv \, \frac{\Phi(v)}{v} \right)- \delta_+ \phi_2(u)\\
&+  \frac{d}{du} \int_0^u d \alpha_1 \int_0^{\bar{u}} d \alpha_3 \frac{1}{\alpha_2} \left( \mathcal{T}_1(\underline{\alpha}) -\frac{1}{2} m_{f_2}^2 \mathcal{T}_2(\underline{\alpha}) \right),
\end{align*}
with 
 \begin{align*}
&\Phi(u)= 3 \phi_\perp(u)+ \delta_+^T\left(\phi_2(u)- \xi \frac{\phi_2'(u)}{2} \right)+\frac{\delta_-^T}{2} \phi_2'(u) \\
&+\frac{d}{du} \int_0^u d \alpha_1  \int_0^{\bar{u}} d \alpha_3 \, \frac{1}{\alpha_2 }\left(\alpha_1 \frac{d}{d \alpha_1}+\alpha_3 \frac{d}{ d \alpha_3} -1 \right) \\
&\left( \mathcal{T}_1(\underline{\alpha}) -\frac{1}{2} m_{f_2}^2 \mathcal{T}_2(\underline{\alpha}) \right).
\end{align*}
For the leading twist DAs we will use the asymptotic form
\begin{align*}
\phi_2(u)=-\phi_\perp(u)=30u \bar{u}(2u-1),
\end{align*}
where we defined $\phi_\perp(u)$ with a minus sign so that we have the same signs in equation (\ref{odd1}) as in Ref. \cite{1007.3541v1} from which we take the value for $f_{f_2}$.

Also using the EOM  \cite{hep-ph/9810475v1} we can express the twist four DAs by the asymptotic form  of lower twist DAs  
\begin{align*}
 g_4(u)&=30 u \bar{u} (2u-1),\\
 h_4(u)&=-30 u \bar{u} (2u-1), \\
  \phi_4(u)&=100 u^2 \bar{u}^2 (2u-1),\\
  \mathbb{A}(u)&=60 u^2 \bar{u}^2(2u-1).
\end{align*}

\bibliography{paperbf2arxiv.bib}

\end{document}